\documentclass[aps,prl,twocolumn,showpacs,superscriptaddress]{revtex4}
\usepackage{amssymb}
\usepackage{amsmath}
\usepackage{graphicx}
\usepackage{hyperref}
\usepackage{epstopdf}
\bibstyle{apsrev4-1}

\begin{document}
\title{Designing hysteresis with dipolar chains}
\author{Andr\'es Concha}
\email[]{aconcha93@gmail.com}
\affiliation{Condensed Matter i-Lab, Diagonal las torres 2640, Building D, Pe\~{n}alolen, Santiago, Chile.}
\affiliation{School of Engineering and Sciences, 
	Adolfo Ib{\'a}{\~n}ez University,
	Santiago, Chile.
}

\author{David Aguayo}
\affiliation{Condensed Matter i-Lab, Diagonal las torres 2640, Building D, Pe\~{n}alolen, Santiago, Chile.}
\affiliation{School of Engineering and Sciences, 
	Adolfo Ib{\'a}{\~n}ez University,
	Santiago, Chile.
	}
\author{Paula Mellado}
\email[]{paula.mellado@uai.cl}

\affiliation{Condensed Matter i-Lab, Diagonal las torres 2640, Building D, Pe\~{n}alolen, Santiago, Chile.}
\affiliation{School of Engineering and Sciences, 
	Adolfo Ib{\'a}{\~n}ez University,
	Santiago, Chile.
	}
\affiliation{Perimeter Institute for Theoretical Physics, Waterloo, Ontario, N2L 2Y5, Canada.}

\begin{abstract}
Materials that have hysteretic response to an external field are essential in modern information storage and processing technologies. A myriad of magnetization curves of several natural and artificial materials have previously been measured and each has found a particular mechanism that accounts for it. However, a phenomenological model that captures all the hysteresis loops and at the same time provides a simple way to design the magnetic response of a material while remaining minimal is missing.  Here, we propose and experimentally demonstrate an elementary method to engineer hysteresis loops in metamaterials built out of dipolar chains. We show that by tuning the interactions of the system and its geometry we can shape the hysteresis loop which allows for the design of the softness of a magnetic material at will.  Additionally, this mechanism allows for the control of the number of loops aimed to realize multiple-valued logic technologies. Our findings pave the way for the rational design of hysteretical responses in a variety of physical systems such as dipolar cold atoms, ferroelectrics, or artificial magnetic lattices, among others.
\end{abstract}

\maketitle
The search for materials with novel magnetic properties has been one of the central subjects in condensed matter physics \cite{handley2000modern}. A large variety of applications have been realized into devices that use these properties at our advantage. Examples of applications are electric transformers, electromagnets, antennas, magnetic resonance imaging, loudspeakers, beam control, mineral separation, and high density memories to name a few \cite{coey2010magnetism}. The specific material to be used in these applications depends on the specific response of it to an external driving magnetic field. In some cases, an irreversible response is needed, and in others a fast reversible response would be preferred. Materials showing an irreversible (hysteretical) response are dubbed hard magnets and soft magnets show lack thereof.

Hysteresis is defined as the irreversibility of a process or the time-based dependence of a system output on present and past inputs \cite{woodward1960particle,handley2000modern}. In physics it is a ubiquitous phenomenon that occurs not only in ferromagnets but also in piezoelectric materials \cite{merz1953double,sluka2013free,xu2015ferroelectric},  in the deformation of soft-metamaterials \cite{florijn2014programmable,lapine2012magnetoelastic}, and in shape-memory alloys to name just a few. The energy dissipated in magnetizing and demagnetizing a magnetic material is proportional to the area of the hysteresis loop. 
Real systems may depict a plethora of hysteresis which include symmetric, asymmetric, rounded, squared or butterfly shaped loops. Due to this shape myriad, the lack of re-traceability of the magnetization curve has been associated not only with domain wall pinning \cite{jiles1986theory}, but also with specific mechanisms of spin rotations (coherent rotation, parallel rotation, fanning, curling) \cite{neel1959d,kittel1949physical,jacobs1957magnetic}, changes of magnetic domains \cite{neel1959d,kittel1949physical,bader2006colloquium},  the nucleation and annihilation of topological defects \cite{guslienko2001magnetization,tchernyshyov2005fractional,zhu2006magnetic,libal2012hysteresis}, and geometrical frustration  \cite{ladak2010direct} among others. 

While each of these mechanisms explains a particular hysteresis loop, it is of utmost interest to find a phenomenological model that captures all these hysteresis loops while remaining minimal. Such a model would allow for the design of the {\it{softness}} of a magnetic material, and for the control of the number of loops aimed to make possible multiple valued logic technologies \cite{dubrova2002multiple}.

\begin{figure}[!t] 
\includegraphics[width=0.50\textwidth,angle=0,clip=true,trim=0 0 0 0]{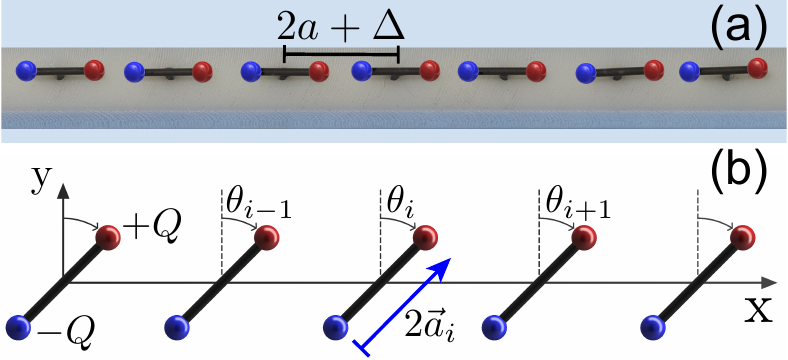}
\caption{($\mathbf{a}$), Experimental realization built out of Neodymium magnets hinged on top of a PTFE plate. Each cylindrical magnet has a length $2a=5\times 10^{-3}$ m, radius 
$r = 0.25\times 10^{-3}$ m, mass $0.007\times 10^{-3}$  Kg, and saturation magnetization $M_{s} =0.92\times 10^{6}$A/m. Their north and south poles are highlighted in red and blue, respectively. The distance between the center of mass of two consecutive rods is $2a+\Delta$.
($\mathbf{b}$), Diagram showing the angular variable of each dipole. Each dipole is modeled as two magnetic charges $+Q,-Q$ at the poles of the rod, where $Q=M_s\pi r^2$. This is know as the dumbbell model \cite{mellado2012macroscopic}.} 
 \label{FIG1}
\end{figure}
 

\begin{figure*}[!th] 
\includegraphics[width=1.0\textwidth,angle=0,clip=true,trim=0 0 0 0]{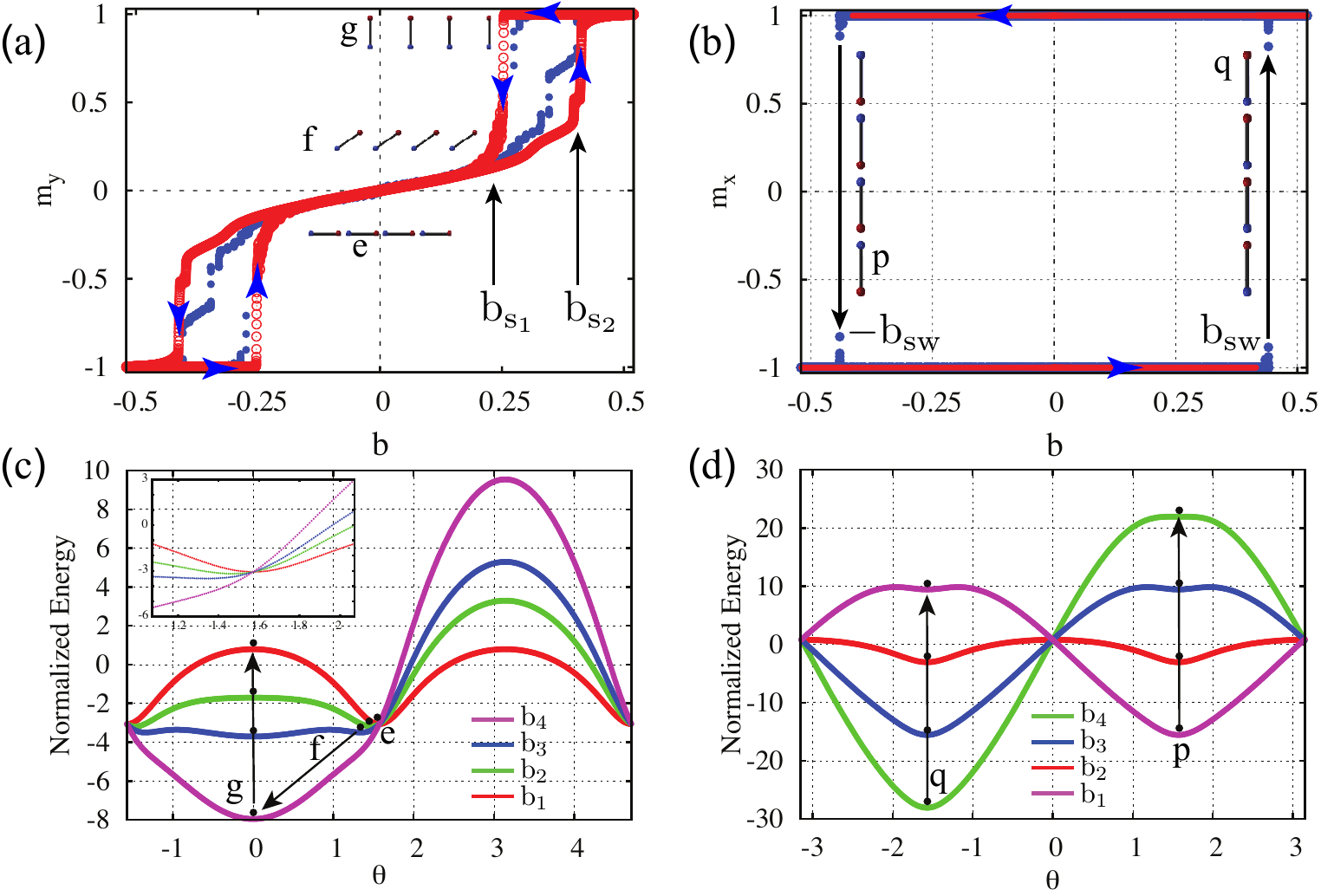}
\caption{($\mathbf{a}$,$\mathbf{b}$) Magnetization loops of a system with $N=10$ (Supplementary Movies 1, 2). Experimental and numerical data in blue dots and red circles respectively. In ($\mathbf{a}$) and ($\mathbf{b}$) the chain is oriented along $\rm{x}$-axis and $\rm{y}$-axis respectively. Blue arrows indicate evolution of the hysteresis cycle. ($\mathbf{c}$) Energy landscapes showing local minima for $\rm{b}_{1}<\rm{b}_{2}<\rm{b}_{3}<\rm{b}_{s2}$ and 
$\rm{b}_{4}>\rm{b}_{s2}$. (e-g) points correspond to the real space configurations shown in ($\mathbf{a}$). Inset shows the energy landscape around point (e). ($\mathbf{d}$) Energy landscapes showing local minima $\rm{b}_{1}<\rm{b}_{2}<\rm{b}_{3}<\rm{b}_{sw}$ and $\rm{b}_{4}>\rm{b}_{sw}$. (p,q) correspond to the real space configurations shown in ($\mathbf{b}$). The energy is normalized by the energy of two interacting magnets with  $\Delta=3$mm.} 
\label{FIG2}
\end{figure*} 
 
Here, we study a magnetic system that displays the main types of reversal loops. The basic unit is a chain of $N$ dipolar rotors  with $\rm{XY}$ rotational symmetry, as shown in Fig.\ref{FIG1}.  Each rotor consists of a hinged Neodymium magnet of length $2a$ and radius $r$ that is free to rotate in the $\rm{XY}$ plane, Fig.\ref{FIG1}b. In our experiments, $1\leq N \leq 12$, rotors were placed at the sites of a Polytetrafluoroethylene (PTFE) plate forming a chain with lattice constant $\Delta+2a$ where $\Delta$  is the shortest distance between the tips of two nearest neighbor rods, as shown in Fig.\ref{FIG1}a. The lowest energy configuration is a head to tail arrangement where two alike magnetic poles stay the closest, Fig.\ref{FIG1}a. This collinear state favors the attraction among magnetics tips of opposite sign. In order to measure the dynamical response of the system, we applied a magnetic field, $\bf{\rm{B}}$, uniform in space and measured the evolution of the rotation angle of each magnet $\theta_i$ (Fig.\ref{FIG1}b). The coarse-grained spin variable for each magnet is defined as $\hat{\rm{m}}_i=(\sin(\theta_i), \cos(\theta_i))$. 

Figure \ref{FIG2}a depicts the experimental (blue dots) and numerical (open red circles) results of a magnetization reversal process of a chain made out of $\textrm{N}=10$ rotors with $\Delta=\frac{6a}{5}$ as a function of $\textrm{b}=B B_{c}^{-1}$ (a dimensionless magnetic field), applied along the $\rm{y}$-axis (Fig.\ref{FIG1}b). $B_c=\left(\frac{\mu_{0}}{4\pi}\right) \frac{Q}{a^2} $ is a characteristic internal field, where  $\mu_{0}$ is the vacuum permeability. 
$\rm{m_y}$ is the average projection of the long axis of the magnets along the $\rm{y}$-axis,  $\rm{m_y}=N^{-1}\sum_{i=1}^{N} \cos\theta_i$. The magnitude of  $\rm{b}$ determines different magnetic configurations and energy landscapes (Fig.\ref{FIG2}a,c):  initially, $\textrm{b}=0$ and the system is in a collinear state,  $\textrm{m}_y=0$ (inset e, Fig.\ref{FIG2}a). As the field grows from zero, magnets are prevented from following the direction of the magnetic field by the internal magnetic interactions. This competition defines the canted phase (inset f, Fig.\ref{FIG2}a). The shallow energy minima defining the canted phase are shown in Fig.\ref{FIG2}c (inset). At $\rm{b}=\rm{b_{s2}}$ they flip from the canted state into a vertical position with average magnetization close to $1$ and remain saturated up to $\rm{b}=b_{\rm{max}}$ (inset g, Fig.\ref{FIG2}a). From saturation, $\rm{b}$ is decreased towards $\rm{-b_{max}}$. The rotors stay parallel until at $\rm{b_{s1}}$  ($\rm{0<b_{s1}<b_{s2}}$) they suddenly rotate towards the $\rm{x}$-axis. As the field decreases, they pass the collinear configuration at zero field and rotate a few degrees as long as $\rm{b}>\rm{-b_{s2}}$, that is, when they suddenly flip toward the $-\rm{y}$-axis  and $\rm{m_y}$ reaches the saturation value $-1$ (inset h, Fig.\ref{FIG2}a). The resulting magnetization reversal (Fig.\ref{FIG2}a) has zero remanence.
Figure \ref{FIG2}b, shows a qualitatively different outcome. Here, $\rm{m_x}$ is the average projection of the magnetization along the $\rm{x}$-axis (Fig.\ref{FIG1}b) and the results (experimental and numerical results represented by blue dots and red open circles respectively) are those of a magnetization reversal when the chain is oriented parallel to the magnetic field. Initially, $\rm{b}=0$ and the system is in a collinear state along the $\rm{x}$-axis, $\rm{m_x}=1$ (inset q, Fig.\ref{FIG2}b). The field decreases from zero towards $\rm{-b_{max}}$ and the system remains still until at the switching field $\rm{b}=-\rm{b_{sw}}$ all rotors suddenly flip towards $-\rm{x}$ resulting in $\rm{m_x}=-1$ (inset p, Fig.\ref{FIG2}b). The process is repeated as the field grows from $\rm{-b_{max}}$ towards $\rm{b_{max}}$, resulting in the rectangular loop shown in Fig.\ref{FIG2}b. The sudden change among metastable magnetic states (Fig.\ref{FIG2}d) corresponds to a first order phase transition induced by the external field. Indeed, the system energy landscape (Fig.\ref{FIG2}d, depicted in red), shows the two collinear states (two minimum at $\theta=\pm \pi/2$) coexisting at  $\rm{b=0}$. The equivalence of these two magnetic states is lifted by the field, and the two possible states become a single one at $\rm{b}=\pm \rm{b_{sw}}$ (green-magenta points Fig.\ref{FIG2}d). \\ 
Next we proceed to examine the dynamics of the system in detail using molecular dynamics simulations. In this description, inertial magnets interact through the full long-range Coulomb potential in the dumbbell approach. 
The equations of motion for the angular variable are:

\begin{equation}
I\frac{d^2\theta_{i}}{dt^2}=\left(\frac{\mu_{0}}{4\pi}\right)\sum_{i \neq j} Q_{i} Q_{j} \vec{a}_{i}\times \frac{\vec{r}_{ij}}{r^{3}_{ij}} -\eta \frac{d\theta_{i}}{dt}+
\left(\vec{P}_{i}\times \vec{B}\right)\cdot \hat{z}
\label{eq1}
\end{equation} 
 where $i$ is the index for charges $\pm Q_{i}$ at the tip of dipole $i$, $\theta_{i}$ is the angle variable shown in Fig.1b main text, and $\vec{a}_{i}$ is the vector that goes from the rotation center to charge $Q_{i}$. In this model, the tip of each rod has a magnetic charge of magnitude $Q=M_s\pi r^2$ that interacts with all other magnetic poles (note that the total magnetic charge per dipole is zero), $I$ is its moment of inertia and $\eta$ is the damping of a rotor in the chain \cite{mellado2012macroscopic}. The last term at the right hand side of Eq.(\ref{eq1}) is the torque due to the action of the external magnetic field on the localized charges at the end of the magnets. $\vec{P}_{i}=2aQ\hat{m}_i$ is the magnetic dipolar moment. We solved this set of coupled equations using the Verlet algorithm.\\
Additionally, energy minimization techniques were used for comparison with the inertial case with damping. We found that hysteresis loops are mainly due to interactions. The set of physical parameters employed in simulations were directly measured by tracking the evolution of a magnet that was slightly perturbed with respect to its equilibrium position (Supplementary Fig. 2, Supplementary Movie 4, and Supplementary Equations 1-3). \\
In this system,  three time scales emerge: 
\begin{eqnarray} 
\tau_{B}=2\pi \left( \frac{I}{2aQB} \right)^{1/2},\hspace*{0.25cm} \tau_{\eta}=I/\eta, \hspace*{0.25cm} \tau_{c}=2 \pi \left( \frac{4\pi a I}{\mu_{0} Q^2} \right)^{1/2}
\end{eqnarray}
They  correspond to the oscillations of a magnet in a uniform magnetic field $B$, the time for the amplitude of its oscillations to decay to $1/e$ its initial value, and the fluctuation time of a neutral plasma of charges $\pm Q$ that are at an initial distance $a$ respectively. The small oscillations of a rotor in a long chain yields the time scale, $\textrm{t}_c \sim I^{1/2} \Delta^{3/2}Q^{-1} a^{-1}$, which is shorter than $\tau_{B}$, $\tau_{\eta}$, and $\tau_{c}$.  Therefore, $\rm{t_c}$ sets an upper bound for the applied field sweeping rate, $\alpha$. 
For the parameters mentioned above ($\Delta=3$ mm) it yields $\textrm{t}_c\sim 4\times 10^{-2}$ seconds. With $\alpha \ll \textrm{b}_{max} B_{c}/ \textrm{t}_{c}$, the magnets have enough time to relax.  Indeed, this is the case as long as the external field sweeping frequency set by $\alpha/(\textrm{b}_{max} B_{c})$, is smaller than the Coulomb frequency $1/\textrm{t}_c$ associated with the fastest dynamical time scale of the system, in which case the width of the loop does not depend on $\rm{\alpha}$.
 
\begin{figure}[!t] 
\hspace*{-0.35cm}\includegraphics[width=0.5\textwidth,angle=0,clip=true,trim=0 0 0 0]{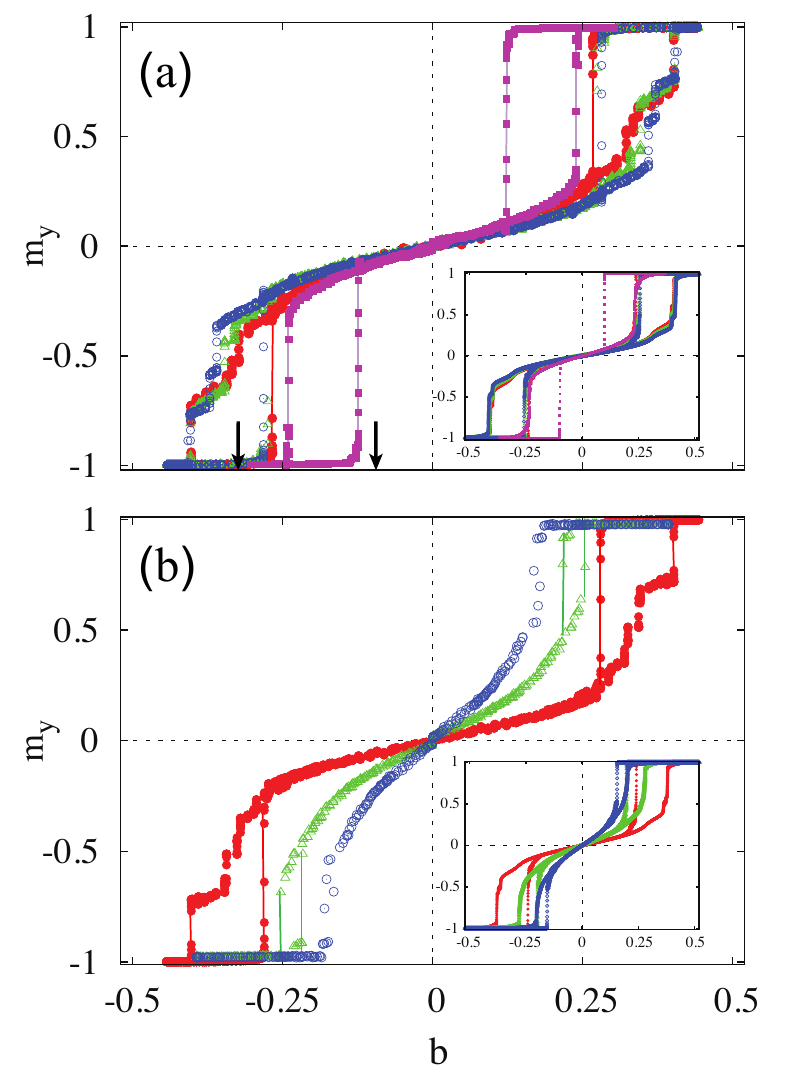}
\caption{($\mathbf{a}$), Experimental hysteresis loops of chains with $N=2$ (magenta  squares), $8$ (red filled circles), $10$ (green triangles) and $12$ (blue open circles) rotors, for $\Delta=6 a/5$. The inset corresponds to hysteresis loops from numerical simulations. Black arrows indicate the positions of the first order estimates for $\rm{b}_{s1}^{N=2}$ and $\rm{b}_{s2}^{N=2}$. ($\mathbf{b}$) Experimental  and numerical (inset), hysteresis loops for chains with $N=10$ and $\Delta= 6 a/5$ (red dots), $8 a/5$ (green open triangles), $2 a$ (blue open circles).}
\label{FIG3}
\end{figure}
\begin{figure}[!t] 
\hspace*{-0.35cm}\includegraphics[width=0.50\textwidth,angle=0,clip=true,trim=0 0 0 0]{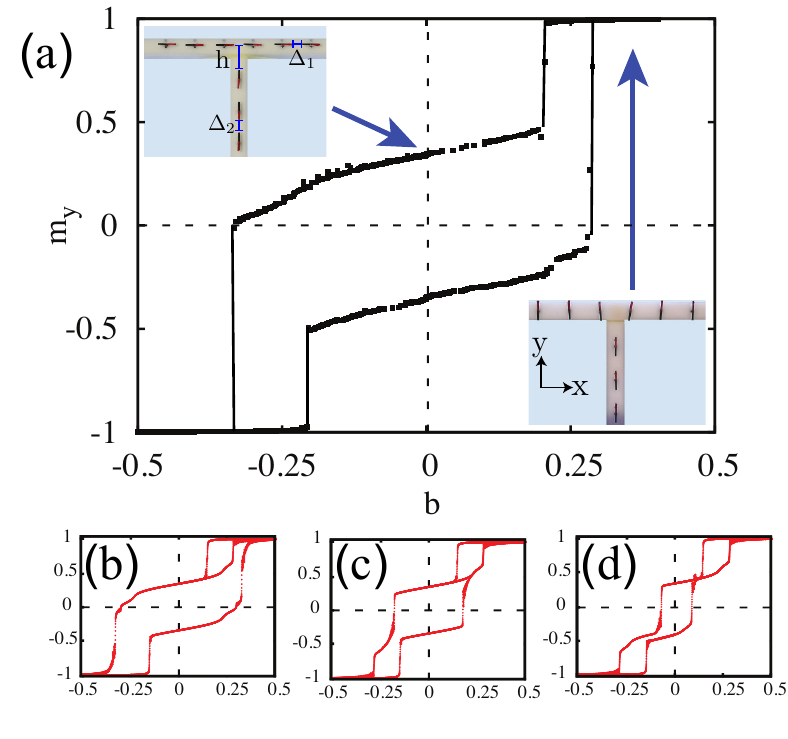}
\caption{($\mathbf{a}$), Hysteresis loop obtained for a system of dipoles made of two linear chains ($\Delta_{1}=3\times 10^{-3}$m,  and $\Delta_{2}=4.0\times 10^{-3}$m) arranged as a T . Real space configurations for initial state and fully polarized state are shown in top, and bottom insets respectively.  The red painted side of each magnet is the magnetic north pole (Supplementary Movie 3). ($\mathbf{b}$-$\mathbf{d}$), show numerical simulations for a T-system when $\Delta_{1}=3.0\times 10^{-3}$m, $\rm{h}=1.5\Delta_{1}$, and $\Delta_{2}/\Delta_{1}=1, 4/3, 2$ respectively. } 
\label{FIG4}
\end{figure}

Having shown how the orientation of the chain respect to the external field determines the shape of the loop at qualitative level, we describe next how the reversal loop of  Fig.\ref{FIG2}a, changes in terms of the physical parameters of the system. In all experiments, $\alpha=3\times 10^{-5}$ T/sec and  $\textrm{b}_{max}=B_{max}/B_c=1.5$, where $B_{c}=2.89\times 10^{-3}$ T. By tuning  $\Delta$ and $N$ we probe how  $\rm{m_y}$ changes with the strength of the coupling among magnets and the size of the system \cite{mellado2012macroscopic}. Figure \ref{FIG3}a shows experimental and numerical (inset) reversal loops for chains with $N=2, 8, 10$ and $12$ and $\Delta= 6a/5$. It is apparent that as $N$ grows, the loops move toward larger fields and quickly saturate. Already the $N=10$ and $N=12$ (green triangles and blue open circles respectively) loops seem very alike. These behavior is in good agreement with numerical simulations (inset Fig.\ref{FIG3}a) .\\
 At a qualitative level, the loop of the $N=2$ chain resembles well the shape of hysteresis loops with larger $N$, as can be seen in Fig.\ref{FIG3}a (Magenta filled squares) (Supplementary Fig.6). Therefore, the $N=2$ case is a useful toy model to compute $\rm{b_{s1}}$ and $\rm{b_{s2}}$ at scaling level. Indeed, 
$\rm{b_{s1}^{N=2}\approx\frac{3}{64\sqrt{2}} \left(4-\Delta/a \right)}$ at leading order in $\Delta/a$ (Supplementary information). Evaluating for $\Delta/a=6/5$ we obtain $\rm{b_{s1}^{N=2}} \approx 0.10$ in close agreement with our experimental results (see Fig.\ref{FIG3}a).
Likewise, a stability analysis at $\theta\sim \pi/2$ (around the collinear configuration depicted in Fig.\ref{FIG1}) yields a formula for $\rm{b_{s2}}$ that allows to quantify the width of the loops in terms of the physical parameters of the magnetic chain and the length of it. Noting that the torque provided by Coulombic forces should equal the one provided by Zeeman interactions $2aQB\sim a \left (\frac{B_{c}a^2 Q}{\Delta^2}\right)$ we can get a simple estimate for 
$\rm{b}_{s2}\sim \frac{a^2}{2\Delta^2}$. Evaluating for $\Delta/a=6/5$ we obtain $\rm{b_{s2}^{N=2}} \approx 0.34$ which is close to our experimental results (see Fig.\ref{FIG3}a).   
\\
Figure \ref{FIG3}b shows the magnetization loops of systems with $N=10$ rotors and $\Delta \in (3, 5)$ mm.  Experiments and simulations (inset Fig.\ref{FIG3}b) show that saturation fields and width of the loops decrease as $\Delta$, increases. Indeed, for larger $\Delta$, the interactions decline and so does it the difference among torques to destabilize parallel and collinear configurations of magnets. From the previous scalings it is clear that even when both saturation fields decrease as $\Delta$ grows, $\rm{b_{s2}}$ does it in a more dramatic fashion, anticipating that as $\Delta$ grows, the width of a hysteresis loop becomes narrower. In the  $\Delta \rightarrow \infty$ limit, magnets decouple from their neighbors and the magnetization reversal approaches that of an isolated rotor. 
It is apparent that the numerical case tends to underestimate saturation fields, this effect being larger for the case of $\rm{b_{s1}}$. In experiments, imperfections in the orientations of the hinges and small variations in the damping coefficient for each rotor favor the departure of all magnets from the parallel state later than in the numerical case. The lower value of $\rm{b_{s2}}$ from numerical simulations is due to the fact that static friction is neglected in the model.
 \\
We further analyze the role of magnetic disorder by studying the dynamics of a chain with magnetic vacancies. Using the same experimental setup, but randomly removing dipoles (Supplementary Figure 9) we find that as the number of vacancies increases, the area of the loops shrink and the concavity of the reversal curve changes as the system approaches the limit of the non interacting case ($N=1$). However, the magnetic response of the basic building block is robust to a low density of vacancies.
\\
Finally, we applied the previous information to design hysteresis loops on demand. In the same spirit as in the Preisach phenomenological model  \cite{woodward1960particle}, the dipolar chains were used as building blocks (hysterons) to produce the complex hysteresis loop shown in Fig.\ref{FIG4} a. This figure corresponds to the magnetization reversal of a T-shaped system (insets in Fig.\ref{FIG4} a) when the external field is applied along the $\rm{y}$-axis. A bigger central loop due to the reversal of the vertical chain is apparent, while the two smaller loops are product of the reversal of the horizontal one. Figs.\ref{FIG4} b-d, show that as the interaction among rotors belonging to the vertical chain decreases ($\Delta_{2}/\Delta_{1}$ grows from $\rm{1.0}$ to $\rm{2.0}$), the three loops (Fig.\ref{FIG4}a) split. Showing that by changing $\Delta_{2}$ it is possible to design the magnetic response using dipolar chains as basic (hysterons) \cite{woodward1960particle,jiles1986theory} building blocks. This demonstrates that we can enhance the coercivity of a system or produce satellite loops if needed. 

The results presented here allow for the design of hysteresis loops with any material whose basic components are interacting polar objects with XY symmetry. There are various examples of such objects  across the scales: magnetic rods or needles at the macroscale \cite{mellado2012macroscopic}, microscopic rotors \cite{fan2005controllable,vacek2007artificial,ketterer2016nanoscale}, magnetic nano-crystals in magnetotactic bacteria \cite{moskowitz1988magnetic}, polar molecules confined in carbon nanotubes \cite{terrones2002molecular,cambre2015asymmetric,ma2017quasiphase}, gases of polar molecules in one dimensional traps \cite{moses2016new}, or individual magnetic atoms \cite{loth2012bistability}. Implementation of logic operations \cite{khajetoorians2011realizing}, multiple valued logic \cite{dubrova2002multiple}, meta-materials with unusual magneto-optical properties \cite{yu2011light,nisoli2013colloquium} or the production of shaped nanoparticles for hyperthermia, a non-invasive technique for drug release or tumor treatment \cite{lee2011exchange} can be realized with current technology using the ideas presented in this letter. It remains an open question if variations of our approach could be used for the study of nontrivial textures such as skyrmions at the macroscale \cite{fert2017magnetic}.

Our proposal is a step forward in the quest for new materials with engineered magnetic properties as it provides an amenable and scalable playground to produce on demand magnetic responses by manipulating interactions, and geometry.  

\begin{acknowledgements}
P.M. acknowledge the Aspen Center for Physics, which is supported by National Science Foundation grant PHY-1066293, the Abdus Salam Centre for theoretical Physics, and Fondecyt Grant No. 1160239. D.A. was supported in part by Conicyt Ph.D. Scholarship No. 21151151.  A.C. was partially supported by Fondecyt grant No. 11130075. The authors acknowledge the support of the Design Engineering Center at UAI. 
\end{acknowledgements}

\end{document}